\documentclass{article}
\usepackage{amsmath,graphicx,style}
\usepackage{amssymb}
\usepackage{booktabs}
\usepackage{mathtools}
\usepackage{textcomp}
\usepackage{enumitem}
\usepackage[colorlinks=true,linkcolor=black,citecolor=black,urlcolor=black,filecolor=black]{hyperref}
\toappear{}

\title{GROUP HETEROGENEITY ASSESSMENT FOR MULTILEVEL MODELS}
\name{Topi Paananen$^\star$, Alejandro Catalina$^\star$, Paul-Christian B\"urkner, and Aki Vehtari\thanks{$\star$ Equal contribution.}}
\address{Helsinki Institute for Information Technology, HIIT\\
Aalto University, Department of Computer Science}
\begin{document}
\maketitle
\begin{abstract}
Many data sets contain an inherent multilevel structure, for example, because of repeated measurements of the same observational units.  Taking this structure into account
is critical for the accuracy and calibration
of any statistical analysis performed on such data.
However, the large number of possible model configurations hinders
the use of multilevel models in practice. In this work, we propose a flexible framework for
efficiently assessing differences between the levels of given grouping variables in the data.
The assessed group heterogeneity is valuable in choosing the relevant
group coefficients to consider in a multilevel model.
Our empirical evaluations demonstrate that the framework can reliably identify
relevant multilevel components in both simulated and real data sets.
\end{abstract}
\section{Introduction}
\label{sec:intro}

Generalized linear multilevel models (GLMMs), also known as hierarchical or mixed models, are a powerful and more flexible extension to the widely popular class of Generalized linear models (GLMs)~\cite{mcculloch2005generalized,gelman2006data}.
The key distinction between the two is that the former models, from a Bayesian perspective, are characterized by a hierarchy of priors and hyperpriors imposed over its parameters that are effectively learned from the data, making them \emph{adaptively} regularised.
This enables modelling of group-structured data by allowing the model parameters to vary across levels of one or more grouping variables.
To be clear on the terminology used, we define the terms we use for referring to group--structured data.
With \emph{grouping} we refer to one way of categorizing the data into different \emph{levels}, and with
level we mean the different instances present in a grouping.

By learning level-specific parameters that deviate from the global population parameters, a GLMM
can learn the differences between levels as well as partially pool their individual parameters at the same time. In particular, partial pooling helps estimating
the parameters of levels with little data.
However, despite their important benefits for many scientific fields and the commonality of group--structured data, GLMMs are still underused in practice where GLMs still dominate.
This can be possibly explained by the GLMMs' higher modelling complexity and required modelling choices: in the absence of domain knowledge
it is difficult to know which predictors have heterogeneous coefficients and in
which groupings.

Ignoring multilevel structure typically results in biased and badly calibrated estimates that fail to fully capture the underlying relationship of the predictors on the outcome.
The three main approaches of naively dealing with group--structured data are:
\begin{itemize}[topsep=0pt,itemsep=2pt,partopsep=4pt, parsep=2pt]
    \item pooling the data of each level to a single point, or
    \item modelling all observations as a single level, or
    \item building separate models for each of the levels.
\end{itemize}
The first approach ignores within--group correlations and dependencies that in most cases are crucial, and often confound the underlying relationships.
The second approach ignores the grouping altogether and fails to learn differences between the levels.
The third approach ignores the similarity of the levels, often ending up overfitting the models in each level, particularly when a level contains few observations.
It is also possible to represent the grouping as one--hot encoded variables, which is closely related to building separate models per group but often computationally inefficient.

In the commonly used \texttt{R}'s formula syntax, an example of a model without group heterogeneity is expressed as \texttt{y \texttildelow \, x1 + x2}, meaning that a response $y$ is determined by two predictors $x_1$ and $x_2$. If we have two group indicators, \texttt{g1} and \texttt{g2}, a model where the regression coefficients are assumed to vary over both group structures is \texttt{y \texttildelow \, x1 + x2 + (x1 + x2 | g1) + (x1 + x2 | g2)}. As the number of predictors and grouping variables increase, there is a combinatorial explosion in the number of possible models.
We propose an approach for quickly assessing which group heterogeneities are big enough so that they should be included in the model. For example, the approach could recommend to construct a GLMM with the syntax
\texttt{y \texttildelow \, x1 + x2 + (x1 | g2)} which is expected to have similar predictive performance as the most complex model.

In this paper, we aim to answer one of the most relevant questions for building a GLMM: which coefficients vary within groups, i.e. across different levels? 
To the best of our knowledge, there are no existing methods that help in answering this question in a robust and automatic manner. %
In this work, we address this issue by proposing a method to perform assessment of group heterogeneity
directly based on the data.
This approach avoids explicitly comparing different models, which
can be time--consuming and prone to selection-induced bias~\cite{cawley2010over,wang2015difficulty}.
By looking specifically at group heterogeneity we aim at telling which coefficients are \emph{different enough} across the levels of a grouping so that they should be taken into account as group coefficients on the target variable.

As the main contribution of this work, we propose a method to assess group heterogeneity
by interpreting the strength of the interaction between numerical predictors and categorical dummy variables as an indicator of group heterogeneity.
We measure this interaction similarly as the interaction of two numerical predictors,
meaning that the joint relationship of two variables with the outcome is different than the
sum of their separate relationships. To perform this with dummy variables, we utilize
the smooth predictions given by
a Gaussian process model, and the KL-diff$^2$ interaction ranking method~\cite{paananen2019ranking}.
We perform extensive experiments showing how the method works in both simulations and real--world data.

The remainder of this article is structured as follows. 
In Section~\ref{sec:previouswork} we discuss previous work addressing similar questions.
In Section~\ref{sec:kldiff} we describe 
our proposed approach for assessing group heterogeneity.
In Section~\ref{sec:experiments} we present both simulations and real world data experiments showing how the method performs in different cases.
Finally, in Section~\ref{sec:conclusion} we summarize the contributions and impact of our work.

\section{Previous Work}
\label{sec:previouswork}

We are not aware of any literature references discussing group heterogeneity assessment explicitly.
Therefore, in this section we discuss methods that are related and, even though they were not designed for this, can be used for this purpose.

As discussed earlier, the interaction of a numerical predictor and a dummy variable can be
interpreted as group heterogeneity if the used model can express these interactions.
In this work, we build on top of the interaction assessment method KL-diff$^2$~\cite{paananen2019ranking}, which we discuss in Section~\ref{sec:kldiff}.
There are similar methods that find pairwise interactions based on the predictions of a model, like H-statistic, and partial dependence values~\cite{friedman2008predictive,greenwell2018simple}. 
These methods have been compared to KL-diff$^2$ and shown to perform worse~\cite{paananen2019ranking}.

Even though this is not a method to evaluate interactions per se, analysis of variance (ANOVA) tables can be used for the purpose of interaction assessment.
However, the modeller first needs to build a full linear model considering already all pairwise interactions, which has a quadratic cost $O(p^2)$. 
Furthermore, in the case of evaluating group heterogeneity, the user needs to know beforehand which grouping factors and predictors to study, which may not be obvious in all cases. 
Because of this, the computational cost is often prohibitive and therefore cannot be regarded as a general method for evaluating interactions.

A recently proposed method finds an equivalent Gaussian process (GP) representation for a given (generalized) linear model to take advantage of the nice properties of the GP's kernel regarding scalability with respect to the number of predictors~\cite{agrawal2019kernel}.
Then, after evaluating the interactions in GP space, they are brought back into the original linear model's space by undoing the transformation.
However, it does not seem to be a general method for finding the equivalent kernel of any linear model. 
Therefore, the usefulness of this method largely depends on the problem and model at hand and thus renders it difficult to use as a general purpose solution.

\section{group heterogeneity assessment}
\label{sec:kldiff}

In this section, we briefly describe the KL-diff method, which uses a Gaussian process
surrogate model to assess the strength of predictor variables and their interactions by differentiating
through the Kullback-Leibler (KL) divergence of the model's
predictive distribution~\cite{paananen2019ranking}. 
The first subsection presents the methodology for assessing the relevance of individual predictors and pairwise interactions.
The second subsection presents our proposal for applying these methods to assess group heterogeneity.

\subsection{KL-diff for ranking predictors and interactions}

After observing data $\mathcal{D} = \{ {X}, {y} \}$, where ${X} \in \mathbb{R}^{N\times D}$ represents the matrix of predictors and ${y} \in \mathbb{R}^N$ the vector of responses, let us denote the posterior predictive distribution of a model $\mathcal{M}$ at some predictor value ${x}^*$ as
\begin{equation*}
p  (y^{*})  \vcentcolon = p  (y^{*} | {\theta}^{*}) \vcentcolon =  p  (y^{*} | {x}^{*},\mathcal{D}, \mathcal{M}) ,
\end{equation*}
where ${\theta}^* = \{ \theta_1^*, \ldots, \theta_{n_p}^* \}$ is the vector of $n_p$ parameters of the distribution that depends on ${x}^*$, $\mathcal{D}$, and $\mathcal{M}$.
We will refer to the posterior predictive distribution corresponding to ${x}^*$ simply as $p(y^*)$.

KL-diff is a local predictive relevance measure for the predictors that takes into account both the mean prediction and its uncertainty through the posterior predictive distribution. 
This concept is formalized as the derivative of a dissimilarity measure, such as Kullback-Leibler
divergence, between two predictive distributions with respect to a predictor.
The method evaluates the derivatives of the KL divergence from the predictive distribution at point ${x}^*$ to the predictive distribution at point ${x}^{**}$ when both points coincide: ${x}^* = {x}^{**}$.
At this point, the first derivative with respect to any predictor $x^*_d$ is zero because the KL divergence has its global minimum of $0$ when both distributions are identical.
Therefore, the authors suggest to use the square root of the second derivative~\cite{paananen2019ranking}.

Based on the predictive distribution that we defined earlier, $p(y^*)$, of a model $\mathcal{M}$, the KL-diff relevance measure with respect to a single predictor variable $x_d$ is defined as
\begin{align*}
    & \text{KL-diff}({x}^*, x_d, \mathcal{M}) = \sqrt{\dfrac{\partial^2 \mathcal{D}_{\text{KL}} [p(y^*)\| p(y^{**})]}{(\partial x^{**}_d)^2}\bigg\rvert_{{x}^{**}={x}^*}} \nonumber\\
    &  = \sqrt{\sum_{k=1}^{n_p}\sum_{l=1}^{n_p}\dfrac{\partial^2 \mathcal{D}_{\text{KL}} [p(y^{*})\| p(y^{*})]}{\partial\theta_k^*\partial\theta_l^*}\dfrac{\partial\theta_k^*}{\partial x_d^*}\dfrac{\partial\theta_l^*}{\partial x_d^*}},
\end{align*}
where $\mathcal{D}_{\text{KL}} [p(y^*) \| p(y^{**})]$ refers to the KL divergence from $p(y^*)$ to $p(y^{**})$ and $\{\theta_1^*, \ldots, \theta_{n_p}^*\}$ are the parameters of $p(y^*)$.
The KL-diff measures with respect to all predictors $x_d$ averaged over the $N$ training
observations can be used as a ranking of their predictive relevance.

If we can use the partial derivatives of the mean prediction with respect to a predictor as a local measure of its relevance, it is natural then that the cross--derivatives of two predictors measures the strength of the joint interaction between both predictors~\cite{paananen2019ranking}.
By computing the cross--derivatives of the Kullback-Leibler divergence between the predictive distributions, an analogous measure of interaction strength is obtained.
Following the same notation as in the previous section, the KL-diff$^2$ with respect to the predictors $x_d$ and $x_e$ is
\begin{align*}
    & \text{KL-diff}^2 ({x^*}, (x_d, x_e), \mathcal{M}) = \sqrt{\dfrac{\partial^4 \mathcal{D}_{\text{KL}} [p(y^*)\| p(y^{*})]}{(\partial x^{*}_d)^2(\partial x_e^{*})^2}} \nonumber\\
    &  \approx \left[ 2 \sum_{k=1}^{n_p}\sum_{l=1}^{n_p}\dfrac{\partial^2 \mathcal{D}_{\text{KL}} [p(y^*)\| p(y^{*})]}{\partial\theta_k^*\partial\theta_l^*} \dfrac{\partial^2\theta_k^*}{\partial x_d^* \partial x_e^*}\dfrac{\partial^2\theta_l^*}{\partial x_d^* \partial x_e^*} \right]^{\frac{1}{2}} .
\end{align*}
The KL-diff$^2$ values represent local measures of pairwise interaction strength of
the predictors $x_d$ and $x_e$.
  
\subsection{Interpreting interactions as group heterogeneity}
\label{sec:multilevel}

Remember that our goal in this work is to identify and assess group heterogeneity in terms of how a given predictor coefficient varies across the levels of a grouping.
The main task is then to identify the predictors whose coefficients have the largest variation between the different levels of one or more grouping factors.
Let us now have some data $\mathcal{D} = \{{X}, {y}\}$ with the coefficients of some of the predictors $x_d$ in ${X}$ varying according to some group structure.
In the simplest setting, with a single grouping variable, we append a dummy vector $g = \{g_1, \ldots, g_N\}$ to the data matrix ${X}$, where $N$ is the number of observations, and where each of the entries $g_i$ is an integer indicating the level to which the $i$th observation belongs.
This formulation naturally generalizes to any number of grouping variables $K$ and denoting individual dummy vectors as $g_k = \{g_{k1}, \ldots, g_{kN}\}$.

With the added dummy vectors, a Gaussian process with a squared exponential covariance function can model complex interactions between the
numerical and dummy variables if they are present in the data.
Then, by applying KL-diff$^2$, we can assess the interaction strength between all $x_d$ predictor variables and all $g_k$ dummy variables to identify potential group heterogeneity.
We note that while it is not our main goal, we can also assess group heterogeneity in terms
of varying intercepts between levels. This can be done by assessing the individual relevances
of the dummy predictors using KL-diff.

The computational cost of running KL-diff$^2$ requires just computing the analytical derivatives of the KL divergence between the predictive distributions.
For Gaussian process models with commonly used likelihoods for numerical, binary or count data, these derivatives
can be computed in analytical form~\cite{paananen2019ranking}.
The method is thus readily applicable to evaluate group heterogeneity for most standard GLMMs.

The proposed methodology is not restricted to Gaussian process models,
but they were chosen because of their several suitable properties.
Firstly, they are able to represent a wide range of linear and nonlinear functions and model high--order interactions.
Secondly, the smoothness properties of the prediction function can be guaranteed and controlled
by the properties of the Gaussian process covariance function.
Thirdly, Gaussian processes can be used in both regression and classification tasks, and
provide well--calibrated predictive uncertainty estimates.

In the Gaussian process covariance function, the predictors are usually assumed to be real-valued.
In Bayesian optimisation, it can be beneficial to modify the Gaussian process covariance function
such that the learned prediction function is discontinuous for integer-valued
variables~\cite{garrido2020dealing}.
However, in this work
we specifically \emph{want} to keep the prediction function continuous even for the integer-valued dummy variables.
That way the model
can learn a smooth prediction function that includes interactions between the numerical predictors and dummy variables arising
from group heterogeneity.

\section{EXPERIMENTS}
\label{sec:experiments}

In this section, we validate the group heterogeneity assessment method discussed in Section~\ref{sec:multilevel} by performing both simulated and real data experiments.
The focus of the experiments is on assessing the performance of the method in terms of correctly identified group coefficients.
However, it is also important to study the false discovery rate, as it can impact the user's budget for model iterations.
Given the lack of alternative methods for directly assessing group heterogeneity, and therefore of baselines as such, we will focus the discussion on the robustness of the method itself (e.g., by studying false versus true discovery rates). In Section~\ref{sec:bikedata} we show that the method finds similar models
as explicitly comparing model candidates using cross-validation.

In all examples, we use squared exponential covariance function in the Gaussian process surrogate model.
For computational reasons, we optimize the hyperparameters of the surrogate model by maximizing the log marginal likelihood, which enables representing the predictive distribution of the Gaussian process in analytical form.
Given the high sample complexity of exact inference in Gaussian processes, this is a commonly used approach that still gives sufficiently well calibrated predictive uncertainty estimates in many cases.

\subsection{Simulated experiments}
\label{sec:simulations}

We perform extensive simulations for regression and binary classification data sets.
For both cases we employ the same data generation process apart from transforming the target $y$.
The predictors $x_d \in {X}$ are normally and independently distributed with mean $0$ and standard deviation $1$.
We sample the population coefficients $b_d$ and intercept $a$ from an uncorrelated multivariate normal distribution.
Additionally, we apply a $0.4$ sparsity factor, so we effectively set $40$\% of the coefficients to be exactly zero. 
Then, each data point is randomly assigned to a level for each group with equal probabilities for the levels.
Like the population coefficients, we also sample the group coefficients from an uncorrelated multivariate Gaussian with group-specific parameters.
The linear predictor is then generated as a sum of both population and group terms. %
The data generation process is summarized below:
\begin{align*}
    x_{id} & \sim \text{Normal}(0, 1) , \, d = 1,\ldots,D, \, i = 1,\ldots,N \\
    b_d & \sim \text{Normal}(\mu_f, \sigma^2_f),\qquad a \sim \text{Normal}(\mu_{b,f}, \sigma^2_{b,f}) \\
    z_d & \sim \text{Bernoulli}(p=0.6) \\
    g_{ik} & \sim \text{DiscreteUniform}(1, L) , \, k = 1,\ldots,K \\ %
    b_{lkd} & \sim \text{Normal}(\mu_{g_{k}}, \sigma^2_{g_{k}}),\qquad a_{lk}  \sim \text{Normal}(\mu_{b,g_{k}}, \sigma^2_{b,g_{k}}) \\
    \mu_i & = a + \sum_{d=1}^D z_db_dx_{id} + \sum_{k=1}^K\sum_{l=1}^L\left[a_{lk} + \sum_{d=1}^D z_db_{lkd}x_{g_k = l,d} \right] \\
    y_i & \sim \pi(f(\mu_i), \phi_i), 
\end{align*}
where $\pi$ is the family's distribution, $f$ is an inverse-link function and $\phi$ the family's specific dispersion parameter, $N$ is the number of observations, $D$ the number of numerical predictors in
the data, $K$ is the number of grouping variables, and $L$ is the number of levels in each group.

For each family, we sample different parameters $\mu_f, \sigma^2_f$ for the population coefficients and $\mu_{g_{k}}, \sigma^2_{g_{k}}$ for the group coefficients, and $\mu_{b,f}, \sigma^2_{b,f}$ for the population intercept and $\mu_{b,g_{k}}, \sigma^2_{b,g_{k}}$ for the group intercepts.
In Table~\ref{tab:family_settings}, we show the chosen family-specific values for these parameters.
They were chosen so that the non-zero coefficients are strong and therefore we mitigate the issues of identifying present but very small coefficients (with true values close to $0$).
For the Bernoulli family, we use the probit link function, and generate the data so that the classes are not clearly separable.

\begin{table}[tbp]
  \centering
  \caption{Values for the family--specific parameters}
  \begin{tabular}[tbp]{rrr}
    \toprule
    Parameter & Gaussian & Bernoulli \\
    \midrule
    $\mu_f$, $\sigma^2_f$ & $5$, $10$ & $0$, $2$ \\
    $\mu_{b,f}$, $\sigma^2_{b,f}$ & $0$, $20$ & $0$, $4$ \\
    $\mu_{g_{k}}$, $\sigma^2_{g_{k}}$ & $0$, $5$ & $0$, $3$ \\
    $\mu_{b,g_{k}}$, $\sigma^2_{b,g_{k}}$ & $0$, $5$ & $0$, $3$ \\
    \bottomrule
  \end{tabular}
  \label{tab:family_settings}
\end{table}

Within this simulation setting, we systematically varied and fully crossed several factors as shown in Table~\ref{tab:settings}.
Within each simulation condition, we repeat the experiments for $50$ data realizations to account for randomness in the data generation process.

\begin{table}[tbp]
  \centering
    \caption{Varied simulation settings}
  \begin{tabular}[tbp]{lr}
    \toprule
    Factor & Values \\
    \midrule
    $N$: Number of observations & $300$ \\
    $D$: Number of predictor variables & $\left\{ 5, 10, 15, 20 \right\}$ \\
    $K$: Number of grouping factors & $\left\{ 1, 2, 3 \right\}$ \\
    $L$: Number of levels in a group & $5$ \\
    $s$: Sparsity  & $0.4$ \\
    $\mathcal{L}$: Likelihood family & Gaussian, Bernoulli \\
    \bottomrule
  \end{tabular}
  \label{tab:settings}
\end{table}

In order to select the relevant heterogeneous group coefficients using the proposed method, we have to apply a decision rule defining which of the returned interactions to choose.
Given that KL-diff$^2$ returns a continuous measure indicating the strength of the interaction, we need to decide which of the returned indicators are strong enough to be chosen.
In this work, we will select the top-$T$ interactions per grouping variable after sorting them, where $T = \lfloor tD\rfloor$, with $t \in [0, 1] \in \mathbb{R}$ and $D$ is the number of possible group coefficients.
In practice, the threshold could be chosen with some selection criterion such as a cross-validation score. 
The selection criterion is left for future research.

\subsubsection{Gaussian simulations}

\begin{figure*}[htb]
\centering
\centerline{\includegraphics[scale=0.8]{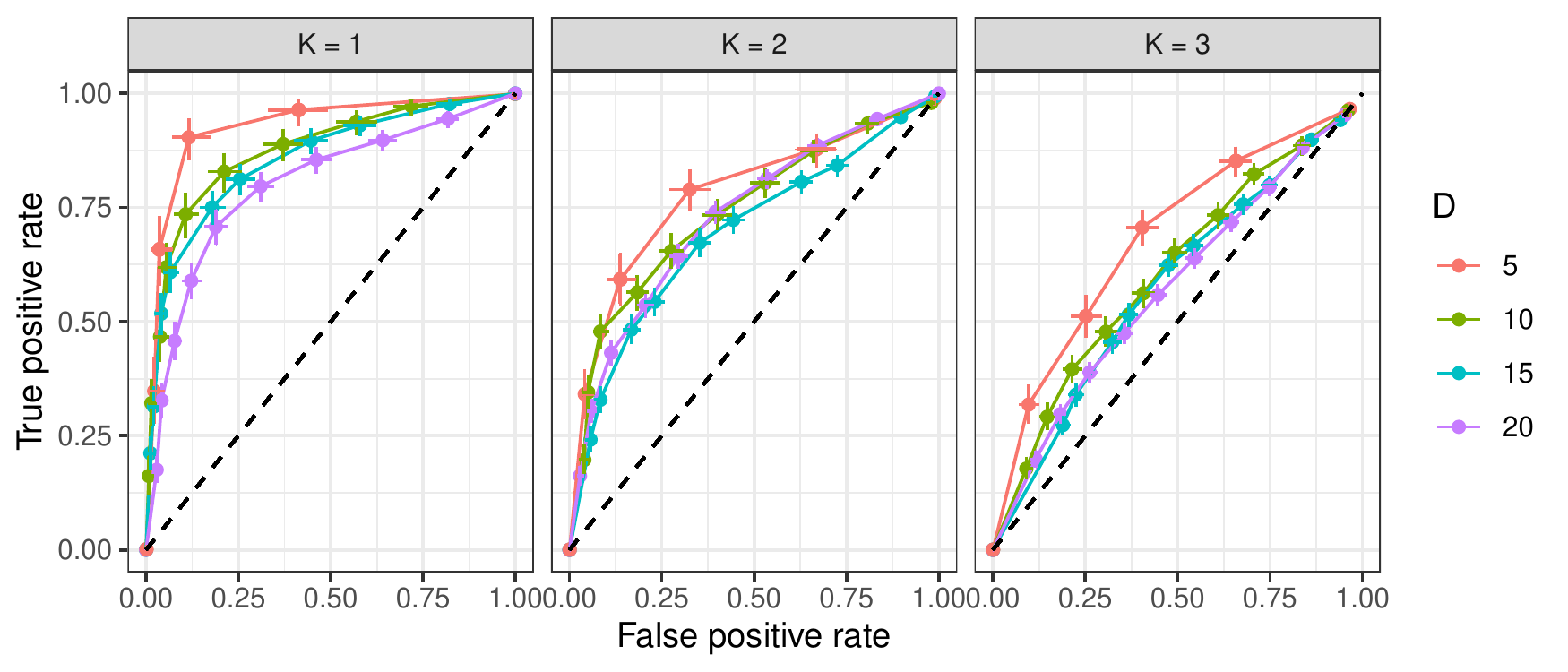}}
\caption{ROC curves for recovered group coefficients as we increase the selection threshold for a model with 40\% sparsity for different numbers of predictors $D$ and grouping variables $K$ in the data. The dashed black line indicates chance selection.}
\label{fig:res_gaussian_t3}
\end{figure*}

Because the Gaussian observation model allows for exact inference under a Gaussian process prior, the predictive distribution has a closed form expression that we use to differentiate following the method described in Section~\ref{sec:kldiff}.

In Figure~\ref{fig:res_gaussian_t3}, we show the results for the Gaussian observation model for different number of grouping variables (as specified in Table~\ref{tab:settings}) and as we increase the interaction selection threshold.
The dots represent the mean estimate and the bars show the $95$\% intervals, both on the y- and on the x-axis.
The figure shows the receiver operating characteristic (ROC) curve where we show the true positive rate as a function of the false positive rate.
For every point in the figure the performance of the method is above chance selection (dashed black line). For just a single grouping variable, 
selection is excellent even with many predictors.
As expected, as we increase the number of grouping variables, the problem gets harder because each
level in the groups explains a smaller part of the total variance.
Given that we have fixed the sparsity in our model to $40\%$, the true number of group coefficients is around $60\%$.
The sparsity affects the performance of the method to some extent, and more thorough analysis of the effect of sparsity is left for further research.
There are several additional factors that may impact the performance of our method.
First, because we are randomly sampling the group coefficients, some of them may end up being too small to be detectable.
Second, our method is not performing feature selection in terms of a \emph{true} model. 
We are evaluating the strength of the interactions through the predictive distribution, and therefore we can only detect group coefficients that make a difference in the predictions.
Because the difficulty depends on the signal-to-noise ratio and the number of observations in the data,
by adding more observations or reducing noise, all the ROC curves get closer to optimal performance.

To keep the experiment concise, the number of levels in each group was held constant. We also did
additional experiments assessing the effect of the number of levels on the method's ability
to detect group heterogeneity. Based on these experiments, having only 2 or 3 levels will make
it difficult to detect heterogeneity, but more than that is enough to detect it.
As for large numbers of levels, even tens of levels did not make the problem more difficult
as long as the number of observations per level is not very small.

\subsubsection{Bernoulli simulations}

In the case of a Bernoulli observational model we no longer have an analytical expression for the predictive distribution as the required integral cannot be solved in closed form.
Likewise, we cannot perform exact inference to find the posterior, and we must turn to approximate inference methods.
In order to use KL-diff$^2$, we use the Laplace approximation to the posterior of Gaussian process latent values.
As before, we optimize the hyperparameters by maximizing the log marginal likelihood, which allows for a closed form solution of the predictive distribution.

Even though we are computing the interaction strengths through an approximate posterior, the results look quite similar to the Gaussian ones.
Due to lack of space and similarity with the previous results, we have omitted the plots of the Bernoulli
results, but the reader can inspect these in the supplementary material located in our github repository
at~\href{https://github.com/topipa/group-heterogeneity-paper}{github.com/topipa/group-heterogeneity-paper}.

\subsubsection{Case study: Usage of rental bicycles} \label{sec:bikedata}

In this section we demonstrate the group heterogeneity assessment framework in a real data set\footnote{http://archive.ics.uci.edu/ml/datasets/bike+sharing+dataset}. The data set
includes the number of daily rental bike uses over two years together with weather information.
The target variable is \texttt{dailyuses}, the number of bike uses per day.
The numerical predictors are temperature, humidity, and wind speed.
We have five grouping variables:
day of the week, month, season, weather category, and a public holiday indicator.
The total data set thus contains  $3$ numerical predictor
variables and $5$ grouping variables, and $731$ observations.
As such, there are already more than $100$ possible predictor-grouping combinations.

We fit a Gaussian process model on the data set and compute the KL-diff$^2$ interaction values between
the numerical and dummy predictors as described earlier.
The eventual goal is to assess the heterogeneity
of the group coefficients for building a GLMM.
To this end, we build
a base model that includes population-level slopes for all the numerical
variables, and varying intercepts for all the grouping variables:
\texttt{dailyuses \texttildelow \, temperature + humidity + windspeed + (1 | month) + (1 | day\_of\_week) + (1 | season) +\\ (1 | weather) + (1 | holiday)}.
Based on the KL-diff$^2$ evaluation, we choose the grouping variable that has the largest total interaction with
the three numerical predictors, which turns out to be the month indicator.
We construct three separate models, each having a single numerical predictor varying over months.
All of the models are fitted using the \texttt{rstanarm} \texttt{R} package~\cite{rstanarm}.

For the three extended models, we
evaluate the group heterogeneity by computing the standard deviation of the posterior means of the slopes between different months, and the predictive performance of the model compared to the base model with leave-one-out cross-validation and the expected log predictive density (elpd) utility~\cite{Vehtari+Gelman+Gabry:2017_practical}.
The results are presented in Table~\ref{tab:bikedata} together with the
KL-diff$^2$ heterogeneity values. The results show that the method
identifies the relevant predictors that have heterogeneous coefficients
between different months, which is present in the resulting
models as both heterogeneity of the slopes between levels, and predictive performance improvement
compared to the base model.
Adding varying coefficients from the other grouping variables does not improve
the predictive performance of the model according to the cross-validation assessment.
Thus, based on the KL-diff$^2$ evaluation, a recommended model formula is
\texttt{dailyuses \texttildelow \, temperature + humidity + windspeed + (temperature + humidity + windspeed | month) + \\ (1 | day\_of\_week) + (1 | season) + \\ (1 | weather) + (1 | holiday)}.

\begin{table}[tbp]
  \centering
  \caption{Group heterogeneity results for the rental bike data}
  \begin{tabular}[tbp]{lrrr}
    \toprule
    Predictor & KL-diff$^2$ & Slope std. & elpd difference\\
    & & between levels & to base model \\
    \midrule
    temperature  & 0.33 & 0.63 & $54 \pm 10$ \\
    humidity     & 0.27 & 0.23 & $36 \pm 9$ \\
    wind speed   & 0.12 & 0.07 & $3 \pm 3$ \\
    \bottomrule
  \end{tabular}
  \label{tab:bikedata}
\end{table}

\section{CONCLUSION}
\label{sec:conclusion}

In this work, we propose a new approach to identify heterogeneous group coefficients from data by using a Gaussian process surrogate model and interaction strength estimates given by the KL-diff$^2$ method~\cite{paananen2019ranking}.
We achieve this by interpreting interactions between numerical predictors and dummy grouping variables as indications for group heterogeneity.

We demonstrate that the proposed method provides a highly useful tool for practitioners who often shy away from hierarchical and multilevel models due to their higher complexity as compared to generalized linear models or other related inference methods.
Quite commonly, the main issue before fitting multilevel models is to assess what the relevant multilevel structure exists in the data (conditioned on the chosen model class).

By tuning the selection threshold, users of the new method gain extra flexibility to either ensure that all the relevant structure is captured or that the resulting model does not exceed their computational resources. 
After identifying the most relevant heterogeneous group coefficients, the user can go and fit the model again explicitly including those coefficients, making multilevel modelling a more accessible option that brings many benefits over standard non--multilevel models.

\section{ACKNOWLEDGEMENTS}

We thank Academy of Finland (grants 298742 and 313122),
Finnish Center for Artificial Intelligence,
and Technology Industries of Finland Centennial Foundation (grant 70007503; Artificial Intelligence for Research and Development)
for support of this research.
We also acknowledge the computational resources provided by the Aalto Science-IT project.

\bibliographystyle{IEEEbib}
\bibliography{multikldiff1}

\end{document}